\documentclass[usenatbib]{mn2e}
\usepackage{graphics}
\usepackage{txfonts}

\title[Q2237+0305 source structure and dimensions]
{Q2237+0305 source structure and dimensions from light curves simulation}
\author [V. G. Vakulik]        
       {V. G. Vakulik$^1$, R.E.Schild$^2$, 
       G.V.Smirnov$^1$, V.N.Dudinov$^1$,  V.S.Tsvetkova$^3$\\ 
       $^1$Institute of Astronomy of Kharkov National University, Sumskaya
           35, 61022 Kharkov, Ukraine\\
       $^2$Center for Astrophysics, 60 Garden Street, Cambridge, MA
           02138, U.S.A.\\
       $^3$Institute of Radio Astronomy of Nat.Ac.Sci. of Ukraine, 
           Krasnoznamennaya 4, 61002 Kharkov, Ukraine\\
           Email: vakulik@astron.kharkov.ua,          
           rschild@cfa.harvard.edu, gleb.smirnov@gmail.com,
           dudinov@astron.kharkov.ua,\\ 
           tsvetkova@astron.kharkov.ua}
        
\date{Accepted ...
      Received ...;
      in original form 2006 August 11}

\pagerange{\pageref{firstpage}--\pageref{lastpage}}
\pubyear{...}

\begin{document}

\maketitle

\label{firstpage}

\begin{abstract}

Assuming a two-component quasar structure model consisting of a central 
compact source and an extended outer feature, we produce microlensing 
simulations for a population of star-like objects in the lens galaxy. 
Such a model is a simplified version of that adopted to explain the 
brightness variations observed in Q0957 (Schild \& Vakulik 2003). 
The microlensing light curves generated for a range of source parameters 
were compared to the light curves obtained in the framework of the OGLE 
program. With a large number of trials we built, in the domain of the 
source structure parameters, probability distributions to find "good" 
realizations of light curves. The values of the source parameters which 
provide the maximum of the joint probability distribution calculated 
for all the image components, have been accepted as estimates for the 
source structure parameters. The results favour the two-component model 
of the quasar brightness structure over a single compact central source 
model, and in general the simulations confirm the Schild-Vakulik model 
that previously described successfully the microlensing and other 
properties of Q0957. Adopting 3300 km/s for the transverse velocity of 
the source, the effective size of the central source was determined to 
be about $2\cdot10^{15}$ cm, and $\varepsilon \approx 2$ was obtained 
for the ratio of the integral luminosity of the outer feature to that 
of the central source. 
\end{abstract}


\section{Introduction}

Because quasar microlensing has the potential to reveal details about the
structure of quasars, large observational data bases at X-ray, optical, 
and even radio wavelengths have been assembled for the Q2237+0305 system 
to compare to theoretical models. The approaches to infer microlensing 
parameters from the light curves of the Q2237 image components may be 
divided into two classes. One of them is based upon the analysis of 
individual microlensing events interpreted as crossing of a caustic fold 
or cusp by the source (e.g., Webster et al. 1991, Shalyapin 2002, Yonehara 
2001, Gil-Merino et al. 2006). The second approach referred hereafter as 
the statistical one, utilizes all the available observational data to infer 
the intrinsic statistical parameters. This approach is represented, e.g., 
by the structure function analysis by Lewis and Irwin (1996), or the 
analysis of distribution of the Q2237 light curve derivatives by Wyithe, 
Webster \& Turner (1999, 2000). Recently, Kochanek (2004) applied a method 
of statistical trials to analyse the well-sampled light curves of Q2237 
obtained in the framework of the OGLE monitoring campaign. 

Both approaches have their intrinsic weak points and advantages. 
In particular, in analysing an individual microlensing event, it is 
necessary to presume that the source actually crosses a single caustic, 
and that the source size is significantly smaller than the Einstein 
radius of typical microlenses. Moreover, there must be some complexity 
caused by the unknown vector difference between the microlens trajectory 
and the macrolens shear. 

In applying the statistical approach, the microlensing parameters are 
obtained through the analysis of the lensed light curves as a whole, 
and much less specific assumptions on the microlensing event peculiarities 
are needed. This approach may encounter the problem of insufficiency of 
statistics, however, and Q2237 is just the case: according to Wambsganss, 
Paczy\'nski \& Schneider (1990) and Webster et al. (1991), light curves 
of duration more than 100 years are needed to obtain reliable statistical 
estimates of microlensing parameters.

The mechanism of accretion onto the massive black hole is presently 
believed to provide the most efficient power supply in AGNs (quasars), 
and effectively all researchers uses various accretion disk models when 
interpreting microlensing events in gravitationally lensed quasars, e.g. 
Rauch \& Blandford (1991), Jaroszy\'nski, Wambsganss and Paczy\'nski 
(1992), and more recent publications by Yonehara (2001), Shalyapin et al. 
(2002), Gil-Merino et al. (2006). However, with the accretion disc being 
generally accepted as a central engine in quasars, the difficulties in 
explaining the observed polarization and spectral properties of quasar 
radiation and their variety still remain, (Ferland \& Rees 1988, Laor 
\& Netzer 1989), as well as the amplitudes of the long-term microlensed 
light curves, which we will discuss in the present paper. 

To explain these discrepancies, some additional structural elements are 
introduced. Outer structural elements inferred are an envelope of 
high-velocity clouds or wind re-emitting the hard chromospheric X-ray 
energy as a network of broad emission lines, and an equatorial torus 
containing the dark clouds that re-absorb the radiation emanating in some 
directions (Antonucci 1993). But the persistent broad blue-shifted 
emission lines previously explained by Antonucci (1993) as an envelope 
of high-velocity clouds have persisted for such a long time in any 
quasar that they must be explained as outflows. Despite many years of 
the idea of outflows, (Chelouche 2005 and references therein), the 
physics of the outflows is not clearly understood as yet. Elvis (2000) 
has suggested the outflow as being launched from the central region 
of quasars and thus, has successfully described how a simple outflow 
structure can explain the emission lines in virtually all quasars. 
The impulse timing analysis of the emission lines by Kaspi et al. (2000) 
for 17 quasars shows that the size scale for these structures must be 
comparable to the size scales implied by auto-correlation analysis for 
continuum radiation by Schild (2005, Fig.1), so it appears that the 
Elvis structures may be expected to originate a significant fraction 
of the quasar's UV-optical luminosity. Direct microlensing measurement 
of the thickness size scale of the emission line emitting region in 
SDSS1004+41 (Richards 2004) implies a size scale comparable to the 
continuum emitting region thickness scale (Schild, Leiter, Robertson 
2006). 

There is observational evidence for the existence of these extended 
structures in the Q2237+0305 quasar. Mid-infrared observations of 
Q2237 made by Agol et al. (2000) favor the existence of a shell of 
hot dust extending between 1 pc and 3 pc from the quasar nucleus and 
intercepting about half of the QSO luminosity. The flux ratios of the 
four Q2237 macroimages measured at $3.6$ cm and $20$ cm by Falco et al. 
(1996) were also interpreted as originating in a source much larger 
than that radiating in the optical wave lengths. Observations in the 
broad emission lines also suggest that they originate in a very large 
structure in Q2237, much larger than that emitting the optical 
continuum, (Lewis et al. 1998, Racine 1992, Saust 1994, Mediavilla et 
al. 1998), though the most recent observations by Wayth et al. (2005) 
indicate a much smaller BEL region, perhaps three times larger than 
the continuum region. We find that the Elvis (2000) outflow model 
easily accommodates these observations.

Microlensing light curves of these complicated source structures may 
noticeably differ from those for a simple source structure represented 
by an accretion disk alone. In particular, the accretion disk alone 
cannot reproduce in simulation the observed amplitudes of the Q2237 
light curves. While providing good fits for the peaks, which are most 
sensitive to the effect of the central source, it fails to provide the 
actual amplitudes of the rest of the light curves, (Jaroszy\'nski et al. 
1992). In this respect, the results by Yonehara (2001), Shalyapin et al. 
(2002), and Gil-Merino et al. (2006) who analysed the regions of the 
light curves near the peaks of HME, provide successful estimates of the 
central source, but ignore the effect of a possible quasar outer feature. 

It should be noted that as early as 1992, Jaroszy\'nski, Wambsganss and 
Paczy\'nski admitted existence of an outer feature of the quasar, that 
reprocesses emission from the disk and may contribute up to 100\% light 
in {\it B} or {\it V}. They simulated microlensed light curves for the 
thin thermal accretion disk model. A bit later, Witt \& Mao (1994) 
demonstrated in their simulations of microlensed light curves of Q2237, 
that a source model consisting of a small central source surrounded by 
a much larger halo structure, would better explain the observed amplitudes 
of the Q2237 light curves. 

Also, the existence of one or several "hot spots" arising in the accretion 
disk was discussed by Gould and Miralda-Escud\'e (1997), and supported  
by Schechter et al. (2003) later, in their analysis of the HE1104-1805 
light curves. Recently, it has been shown how the microlensing light curves 
may be affected by a presumed fractal structure in the X-ray emitting 
region  (Lewis 2004), and in the broad line region (Lewis \& Ibata 2006).  
In 2003, Schild \& Vakulik have shown how the double-ring model of the 
Q0957 surface brightness distribution, resulting from the Elvis (2000) 
quasar spatial structure model, successfully explains the rapid 
low-amplitude brightness fluctuations in Q0957+561. Reference to 
microlensing models of Schild and Vakulik (2003) allowed inferences about 
structure sizes, and Schild (2005) even showed that the orientation of 
the quasar on the plane of the sky can be determined. Interestingly, 
Abajas et al. (2002, 2007) have recently simulated the emission line and 
continuum light curves produced by microlensing of a bi-conic outflow 
region for a variety of the bicone orientations with respect to the shear, 
direction of motion and sight line.
 
The standard accretion disc model also cannot explain the large color
effects associated with microlensing discovered by Vakulik et al. 
(2004). Color variations in microlensing of an accretion disc with a 
radial temperature-color gradient have been predicted by Kayser, Refsdal 
\& Stabell (1986) and simulated later by Wambsganss and Paczy\'nski 
(1991). An excellent and careful simulation by Jaroszy\'nski, Wambsganss, 
\& Paczy\'nski (1992) showed that microlensing color effects comparable 
to those observed are possible with their classical geometrically thin, 
optically thick accretion disc model, but they predicted a rather small 
source and too large brightness fluctuations in the simulated light curves. 

In the sections to follow, we analyse the Q2237 microlensing light 
curves using a two-component model of the quasar's structure, and apply 
a statistical approach to determine parameters of this two-component 
source model. The approach we applied is in general similar to that of 
Kochanek (2004). In contrast to our ring model proposed earlier (Schild 
\& Vakulik 2003), we used a simplified model, consisting of a compact 
central source and an extended outer structure with a much smaller 
surface brightness. Such a model, being much easier for calculations, 
possesses the principal property of the ring model to produce sharp 
peaks of the simulated light curves, while damping the amplitudes of 
the entire microlensing event light curve. 

Thus, our basic approach is to accept the existence of inner and outer 
structural elements as detailed above, and to derive from parameter 
fitting only the size of the inner luminous feature, and the fraction 
of the total UV-optical energy from the extended outer feature as 
compared to the luminosity originating in the compact central feature. 
We will show that the structural elements of this two-component quasar 
model satisfactorily explain the observed microlensing brightness curves.

\section{Simulation of light curves}

In the vicinity of a selected macroimage, the principal lens equation 
can be represented in linearized form as proposed by Kayser, Refsdal \& 
Stabell (1986) and Paczy\'nski (1986):
\begin{equation}
\vec y(\vec x)=\vec x-{\left(\begin{array}{cc}\sigma_*+\sigma_c+\gamma&0\\
0&\sigma_*+\sigma_c-\gamma\end{array}\right)}\ \vec x
\end{equation}
where $\vec x$ is a dimensionless coordinate in the lens plane, and 
$\vec y$ is a corresponding coordinate in the source plane, $\sigma_*$ 
is the normalized surface density of microlenses (stars), and $\sigma_c$ 
is a surface density of any diffusely distributed matter. Parameter 
$\gamma$ (shear) characterizes asymmetry of the gravitational field 
distribution in the region under consideration. In numerical simulation, 
the stellar constituent $\sigma_*$ of the normalized surface density 
$\sigma$ can be represented in an explicit form by an ensemble of 
microlenses randomly distributed in the macrolens plane, and then the 
previous formula can be written in the form:
\begin{equation}
\vec y(\vec x)=\vec x-{\sum_im_i{{\vec x-\vec x_i}\over{|\vec x-
\vec x_i|^2}}}-{\left(\begin{array}{cc}
\sigma_c+\gamma&0\\0&\sigma_c-\gamma\end{array}\right)}\vec x
\end{equation}

Here, $\vec x(x_1,x_2)$ and $\vec y(y_1,y_2)$ are the beam coordinates 
in the lens plane and in the source plane, respectively, $\vec x_i$  
is the coordinate of the {\it i}-th microlens (a star), and $m_i$ is 
the  microlens  mass expressed in the units of the solar mass. All 
coordinates are dimensionless in this equation and are given in units 
of the Einstein radius of a single star, in the source plane and lens 
plane, respectively:
\begin{equation}
r_E^2={{4Gm}\over{c^2}}{{D_SD_{LS}}\over{D_L}}, \quad R_E^2={{D_L^2}
\over{D^2_S}}{r_E^2}.
\end{equation}
Here, $G$ is the gravitational constant, $c$ is the velocity of light in 
the vacuum, $m$ is a microlens mass, and $D_{LS}$, $D_L$, $D_S$ are 
cosmological distances between the lens, the source, and the observer, 
(Schneider et al. 1992, Witt \& Mao 1994).

Using equation (2), with the ray tracing method, (Schneider et al. 1992), 
it is possible to calculate the distribution of magnification rate 
$M(y_1,y_2)$ for a small (quasi-point) source for all locations $(y_1, 
y_2)$ -- the so-called magnification map. In microlensing of a finite-size 
source, which is situated at the $(y_1^\prime, y_2^\prime)$ point in the 
magnification map and is characterized by a surface brightness distribution 
$B(y_1,y_2)$, the magnification rate can be calculated from the formula:
\begin{equation}
\mu(y_1^\prime, y_2^\prime)={{\int\int B(y_1, y_2)M(y_1-y_1^\prime,
y_2-y_2^\prime) dy_1dy_2}\over {\int\int B(y_1, y_2)dy_1dy_2}},
\end{equation}
where the integrals are calculated within a region where the surface 
brightness $B(y_1,y_2)$ is non-zero. In this expression, we use the 
values of $M(y_1,y_2)$ from the magnification map calculated for a 
quasi-point source.
Specifying the source trajectory at the magnification map by the 
projections of its velocity $V_1, V_2$ at the coordinate axis, one can 
calculate the corresponding simulated light curve: 
\begin{equation}
m(t)=-2.5\lg[\mu(y_{01}+V_1t, y_{02}+V_2t)]+C
\end{equation}
 where $y_{01}, y_{02}$ are selected starting points of the trajectory.

One of the principal advantages of the ray tracing method is that, 
once calculated, the magnification map can be used to produce a large 
set of simulated light curves for various models of surface light 
distribution $B(y_1,y_2)$ of a lensed source.

We simulated microlensing of a two-structure source, with one of them 
the compact, central luminous source having a surface brightness 
distribution $B_1(y_1,y_2)$.  The other, outer, structure, is associated 
with the larger structural elements -- a shell, a torus, Elvis's 
biconics (Elvis 2000) -- and is characterized by substantially lower 
surface brightness $B_2(y_1, y_2)$. For such a source, situated at point 
$(y^\prime_1,y^\prime_2)$, the values of the magnification rate 
$\mu_{12}$ can be calculated from:
\begin{equation}
\mu_{12}(y_1^\prime,y_2^\prime)={{\mu_1(y^\prime_1,y^\prime_2)+
\varepsilon\mu_2(y_1^\prime,y_2^\prime)}\over{1+\varepsilon}}.
\end{equation}
The magnification rates $\mu_1$ and $\mu_2$ are calculated according to
(4) for the surface brightness distributions $B_1$ and $B_2$, while 
$\varepsilon$ is determined as a ratio of the integral luminosities of 
these structures:
\begin{equation}
\varepsilon={{\int\int B_2(y_1,y_2)dy_1dy_2}\over{\int\int B_1(y_1,y_2)
dy_1dy_2}}.
\end{equation}

The characteristic time-scale of observed Q2237 microlensing brightness 
fluctuations is known to be almost a year. We infer from known cosmological
transverse velocities that such a scale is due to microlensing of the 
compact inner quasar structure. Since the predicted spatial scale of the 
outer structure is more than an order of magnitude larger as compared to 
the inner part, (Elvis 2000, Schild \& Vakulik 2003), the expected time 
scale of its microlensing brightness variations should exceed ten years. 
So, because of the 
large dimension, the amplitudes of microlensing magnification should be 
noticeably less, as compared to microlensing of the compact structure.
Thus we conclude, that on time-scales near 4 years, the magnification 
rate $\mu_2(y_1,y_2)$ is almost unchanging and does not differ noticeably 
from the average magnification rate of the {\it j}-th component $\mu_j$ 
resulting from microlensing: $\mu_2(y_1,y_2)\approx \langle\mu_2(y_1,y_2)
\rangle\approx \mu_j.$ Under these assumptions, equation (6) can be 
rewritten:
\begin{equation}
\mu_{12}(y_1^\prime,y_2^\prime)={{\mu_1(y^\prime_1,y^\prime_2)+
\varepsilon\mu_j}\over{1+\varepsilon}}.
\end{equation}

It is clear that, under such assumptions, microlensing of the extended 
(outer) structure does not produce noticeable variations of magnification 
or brightness fluctuations on the observationally sampled time-scales, 
and therefore is effectively a brightness plateau above which the inner
structure brightness fluctuations are seen. So the observed inner region
brightness fluctuations are reduced by $1/(1+\varepsilon)$. 

Therefore, when analysing the light curves for 4 year time intervals, 
we did not attempt to estimate the size of the extended structure, and 
the accepted value of $\varepsilon$ was the only parameter which 
characterized the outer structure. The inner compact structure of the 
source was simulated by a disc with a Gaussian surface brightness 
distribution, and with its characteristic size $r/r_E$ at the one-sigma 
level as a fitted parameter. We used this simple central source model 
because it is more easy for computation. In doing so, we relied on the 
work by Mortonson et al. (2005), who examined the effect of the source 
brightness profile on the observed magnitude fluctuations in 
microlensing. They used a variety of accretion disc models, including 
Gaussian disk, and concluded that the statistics of microlensing 
fluctuations is relatively insensitive to a particular light 
distribution over the source disk excepting the effective radius.

 \begin{figure}
 \resizebox{\hsize}{4.5cm}{\includegraphics{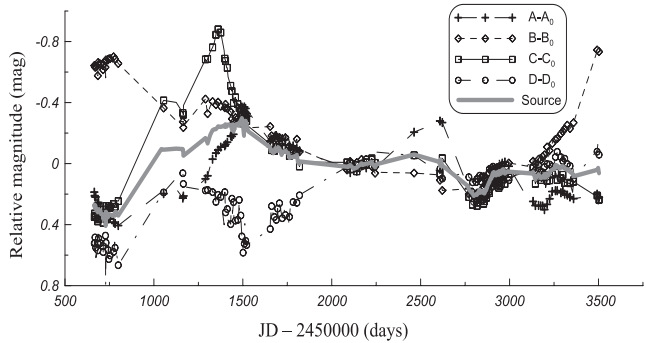}}
 \caption{The OGLE light curves obtained in the {\it V} filter in 
 1997-2005.  The light curves  are reduced to the same (zero) level 
 at the time interval JD 2090-2250. The solid grey line is our 
 estimate of variations of the source brightness.}
 \label{fig1}
 \end{figure}
 
 In producing magnification maps, microparameters $\sigma_*$ and $\gamma$ 
(shear) were taken from Kochanek (2004): 0.392, 0.375, 0.743, 
and 0.635 for  $\sigma_*$, and 0.395, 0.390, 0.733, 0.623 for $\gamma$, 
for the A, B, C and D components, respectively. For each of the four 
Q2237+0305 components, we calculated five magnification maps with dimensions 
of 30 x 30 microlens Einstein radii, (a pixel scale of $0.02\ r_E$).
We assumed here that the 
entire mass is concentrated only in stars -- that is, $\sigma_c$ equals zero. 
To simplify computations, all the stars were modeled as having the 
same mass. 
We are well aware that this assumption is rather artificial, but in doing so,
we refer to the works by Wambsganss (1992) and Lewis \& Irwin (1995) who 
demonstrated, having used various mass functions, that the resulting 
magnification probability distributions are independent of the mass function 
of the compact lensing objects. For the sake of truth, the more recent works
by Schechter, Wambsganss \& Lewis (2004) and Lewis \& Gil-Merino (2006) 
should be mentioned, where simulations with two populations of microlenses 
with noticeably differing masses were carried out to show that the 
magnification probability distributions can depend on the mass function. It 
is not surprising that such an exotic case has demonstrated the effect the 
authors wanted to demonstrate. But it is of little relevance for our work, 
since the more relevant calculation of Lewis \& Irwin (1995) was made for 
more realistic mass functions -- for microlenses in $0.3M_\odot$,  
$10.0M_\odot$ stars, and for the Salpeter mass function; these do not 
show appreciable dependence of the magnification probability distribution 
on the adopted mass function.

\section{Simulation procedure and results} 

For our analysis, we used the Q2237+0305 light curves obtained in the 
{\it V} filter by the OGLE group in 1997-2000. High sampling rate 
(2-3 datapoints weekly) and low random errors are inherent in the 
photometric data from this program. To compare with the simulated light 
curves, results of the OGLE photometry were averaged within a night, thus 
providing 108 data points in the light curve of each image component.
For every set of the source model parameters, we estimated the 
probability to produce close approximations to the observed 
brightness curves. The values of the source model parameters 
providing the maximum probabilities, were accepted as the 
parameter estimates. To estimate consistency of the results, 
the analysis was carried out for each of the four image components 
separately.

Quasars are known to be variable objects, and  their luminosity
may change noticeably on time-scales of several years, months, 
and even days, (De Vries 2005, Gopal-Krishna et al. 
2003, Rabbette et al. 1998). If a variable source 
is macrolensed, the intrinsic brightness variations will be observed 
in each lensed image with some time delays. This is just the fact 
that allows the Hubble constant to be
determined from measurement of the time delays.
In analysing microlensing, however, variability of the source is an 
interfering factor, which needs to be taken into account. In the Q2237+0305 
system, because of an extreme proximity of the lensing galaxy, ($z=0.04$),
and because of the almost symmetric locations of the 
macroimages with respect to 
the lens galaxy center, the expected time delays do not exceed a day, 
(e.g. Wambsganss \& Paczy\'nski 1994, Schmidt et al. 1998). 
This is why the intrinsic brightness variations of the source would 
reveal themselves as almost synchronous variations of brightnesses of all 
the four lensed images. This was observed in 2003, (Vakulik et al. 
2006), when the microlensing activity was substantially subdued 
for all four image components.
 
Generally, separation of the intrinsic brightness variations of the source 
from the light curves  containing microlensing events is a poorly defined 
and intricate task.
In our attempts to obtain the intrinsic source light curve for Q2237+0305, 
we introduced the following assumptions:

1. No effects of microlensing on the brightnesses of the components were 
observed during the time interval from January to June, 2002, (JD 2090-2250), 
when the magnitudes of all the components were almost unchanged. The magnitude 
of each component was accepted as a zero level, and its brightness variations
were analysed relative to this level.

2. Relative to this zero level, we regarded that the closer these 
light curves were to each other, the higher the probability that the 
components are not microlensed within this time interval, while 
almost synchronous variations of their brightness are due to changes of 
the quasar brightness. And vice versa, the more the light curve of 
a component deviates from others, the larger the probability that the 
component is subjected to microlensing, which veils and distort the 
source variations.

Thus, the weighted average variations of the component brightnesses 
with respect to their zero levels were adopted as the estimate 
of the source brightness curve, with the statistical weight for 
variations of every component being selected depending on how close 
this brightness variation to variations of other components is. The 
quasar intrinsic brightness curve, obtained on the basis of 
these assumptions, as well as the OGLE light curves for the individual 
images, reduced to their zero levels, are shown in Fig. 1. 
We expect the largest source of error in this intrinsic source 
brightness history to be encountered during the time interval from August 
1998 to December 1999, when, possibly, all the components underwent 
microlensing.
 \begin{figure}
 \resizebox{\hsize}{4.5cm}{\includegraphics{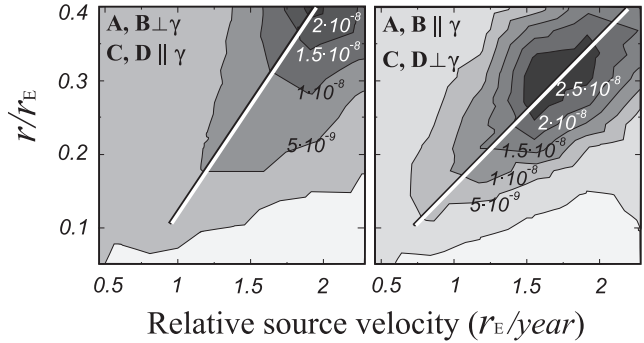}}
 \caption{Probability distributions to find "good" simulated light curves 
 as  functions of the scaling factor and relative dimension of the compact 
 feature.  The diagrams are built for all the four components of 
 Q2237+0305 for two  directions  of the source motion (as indicated on 
 both panels in the left  bottom corner). The probability scale is shown 
 in levels of grey.}
 \label{fig2}
\end{figure}

To characterize similarity of the simulated and observed light curves, 
 a  $\chi^2$ statistics for each image component was chosen:
\begin{equation}
\chi^2_j=\sum_j^{N_S}{{[m_j(t_i)-M_j(t_i,\vec p)]^2}\over{\sigma^2_j}},
\end{equation}
where $m_j(t_i)$ is the observed light curve of the $j$-th component, and 
$M_j(t_i,\vec p)$ is one of the simulated light curves, produced from a 
 source trajectory at the magnification map, and $N_S$ is a number of points in the
 observed light curve. The 
magnification map was calculated for the source model described by a set 
of parameters $\vec p$, which could be varied. The quantity 
$\sigma^2_j$ characterizes the errors of the observed light curve 
measurements, which are $0.032^m$, $0.039^m$, and $0.038^m$  
for the A, B and C components, and $0.057^m$ for the faintest D component. 

The probability that, for a given set of parameters $\vec p$, a simulated 
light curve will be close enough to the observed light 
curve, -- that is, the value of $\chi^2$ will happen to be less than some 
boundary value $\chi^2_0$, -- such a probability will be:
\begin{equation}
P({\chi^2<\chi_0^2})={{N_{\chi^2<\chi^2_0}}\over N_{tot}},
\end{equation}
where $N_{\chi^2<\chi^2_0}$ is a number of successful trials, and $N_{tot}$
is a total number of trials. The boundary value $\chi^2_0/N_S=3$ was 
adopted for calculations.

The direction of an image motion with respect to the shear is an 
important parameter, which affects the probabilities noticeably. 
In (Kochanek 2004), directions of motion of each component 
were chosen randomly and independently of directions of other 
components. This is not quite correct, since the directions of 
motion of components are not independent, and are determined by 
the motion of the source. Therefore, specifying the motion of 
one of the components must automatically specify motions of other 
components, if the bulk velocities and velocity dispersion of 
microlenses can be neglected (Wambsganss \& Kundi\'c 1995, Wyithe 
et al. 2000). As a result of almost perfect symmetry of Q2237+0305, 
for any direction of the source motion, directions of the opposite 
components with respect to the shear direction must coincide, while 
motion for the two other images must be perpendicular to the shear 
direction. That's why, unlike the work by Kochanek (2004), we 
analysed trajectories for two selected directions at the 
magnification maps, -- when the A and B components are moving 
along the shear, with the C and D moving transversely to it, 
and vice versa, when A and B are moving transversely to the shear.
Also, we did not undertake the local optimization of trajectories, 
as in Kochanek (2004), since this may distort 
the estimates of probabilities.

Thus, for each magnification map, and for each of the two selected 
directions, a map of the distribution of the initial points of the 
trajectories can be calculated, for which $\chi^2<\chi^2_0$. 
The probability (10) can then be calculated as the relative area of 
such regions on the map.

To reduce computing time, the map of $\chi^2$ was calculated 
initially with a coarse mesh, ($\sim 0.3\, r/r_E$), to localize the 
regions with low values of $\chi^2$. Then, more detailed calculations 
with a finer mesh were carried out for only these regions.

In our simulation, the following parameters could be varied: the radius of 
the central compact feature $r$, expressed in units of a microlens Einstein 
radius $r_E$; the brightness ratio, $\varepsilon$, expressing the ratio of 
the total outer structure (Elvis structure) brightness to the total inner
structure brightness; and the relative transverse velocity of the source, 
$V_t (r_E/year)$, expressed in the units of the Einsten radius of a 
microlens per year, which is also a scaling factor for simulated 
light curves. 

 \begin{figure}
 \resizebox{\hsize}{18cm}{\includegraphics{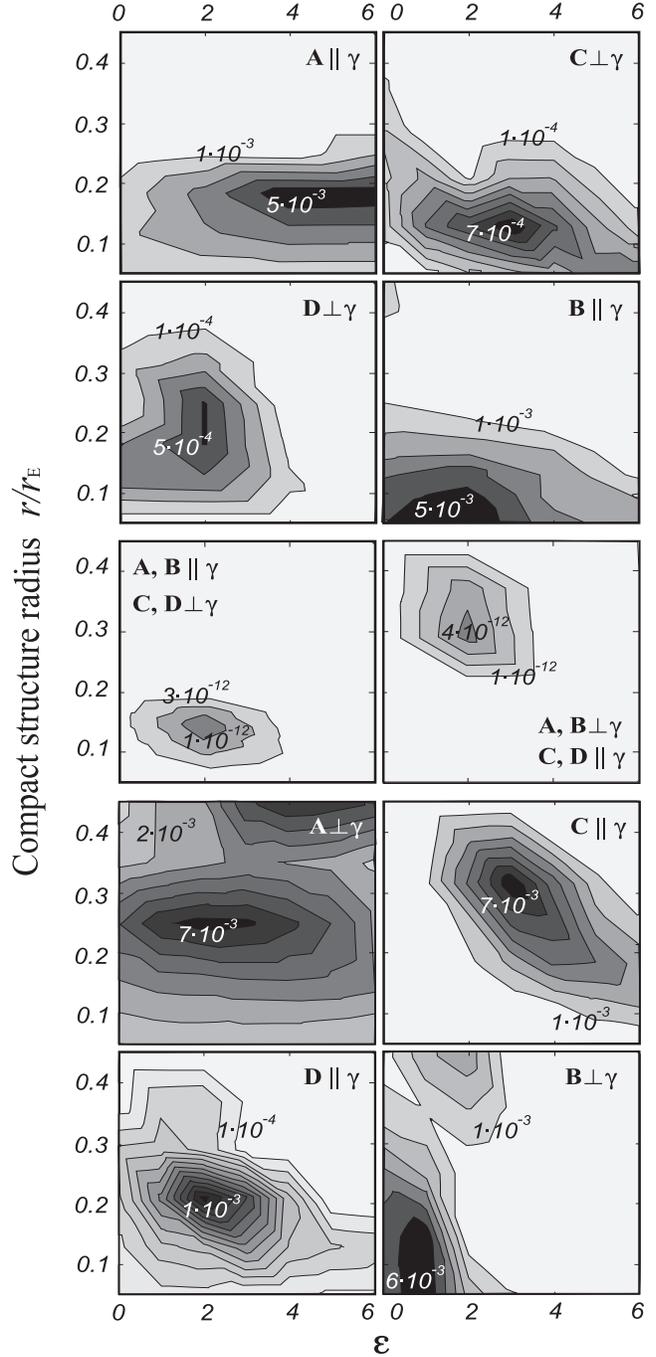}}
 \caption{Probability distributions to find "good" simulated light curves 
 as  functions of the compact structure dimension, $r/r_E$, and of the 
 ratio  $\varepsilon$  of the integral brightnesses of the outer and inner 
 source  structures. Rows 1 and 2: images C and D move along the shear, 
 A and B  transversely. Rows 4 and 5: A and B  move parallel to the shear, 
 C and D -- transversely. The panels in the 3-d row show the joint
 probability distributions for the four components, for the two directions 
 of the source motion.}
 \label{fig3}
\end{figure}

The search of probabilities for our three fitting parameters, $r/r_E$, 
$\varepsilon$ and $V_t$ is a rather complicated  task, which needs much 
computing time. We attempted to simplify it in the following way. 
Assuming that the effect of the outer structure on the characteristic 
time scales of microlensing brightness variations is insignificant, we 
put $\varepsilon=0$ at the first stage. Hence, a dependence of 
probabilities on two parameters, -- the scaling factor and relative 
dimension of the compact feature, -- was evaluated at the first stage. 
In Fig.2, the diagrams are presented, which demonstrate distributions 
of probabilities to find simulated light curves, which would be close 
to the observed ones, -- depending on the scaling factor and the compact 
feature dimension. The diagrams are built for all the four components 
for two directions of the source motion: A and B along the shear, C and 
D transversely -- at the left, and C and D along the shear, A and B 
transversely -- at the right. Overall, a nearly linear dependence of 
the scaling factor from the source dimension is seen. The largest values 
of probability occur for $r/r_E \approx0.4$ in the first case, and for 
$r/r_E\approx0.3$ in the second case. Since the maximum of probability 
distribution is, on average, at the source radius of $0.35\ r_E$ and 
the velocity of $1.82\ r_E$ per year (see Fig. 2), and taking into 
account a nearly linear dependence between these values, we adopted the 
following expression for the further calculations: 

\begin{equation}
V_t=1.82^{+1.18}_{-0.52}\ {{r/r_E}\over{0.35}}.
\end{equation}
Here, the source velocity $V_t$ is expressed in the units 
of the microlens Einstein radius per year.

At the second stage of our simulation, the distribution of the 
probabilities to find simulated light curves similar to the observed 
ones was estimated, depending on the source's compact feature dimension 
$r/r_E$, and on the relative integral brightness of the outer feature 
$\varepsilon$. The scaling factor $V_t$ was determined according to 
(11) for each determination of the source dimension. The diagrams 
constructed for two different directions of the source motion, -- along 
the shear $\gamma$  and transversely to it  for each component, -- are 
shown in Fig. 3. Joint probability distributions for all the four 
components, calculated as $P_{all}=P(A)P(B)P(C)P(D)$, are also shown 
in the third row in this figure.

It is very significant that the probability maxima for all the components 
are found for values of $\varepsilon$ larger than zero. This means that 
the outer quasar structures must noticeably contribute to the total 
quasar brightness in the optical wavelengths. The contribution decreases 
the amplitudes of microlensing brightness fluctuations and is at the 
core of the conundrum that in Q2237, observed microlensing events  
are lower in amplitude than inferred for simple luminous accretion disc 
models. 

Interestingly, we also see from Fig. 3 that the statistics and 
locations of probability maxima in the domain of parameters 
$\varepsilon$ and $r/r_E$ found for each of the components separately, 
differ for different directions of motion of the image components 
with respect to the lens shear, $\gamma$. Examining four upper 
and four bottom panels of our Fig. 3, we see that the values of 
probabilities for the A and B components at their maxima are almost 
the same for the two selected directions of the source motion, while 
the C and D components both exhibit higher probabilities for the 
source to move parallel to the line connecting C and D rather than A 
and B. However, the joint probabilities calculated for all the four 
components, (the third row panels of Fig. 3), though giving slightly 
differing values of $\varepsilon$ and $r/r_E$, favor neither of these 
two cases in terms of the maximal values of probabilities. Much 
larger statistics is needed to solve this important problem, which 
is beyond our current computational resources.
 \begin{figure}
 \resizebox{\hsize}{6cm}{\includegraphics{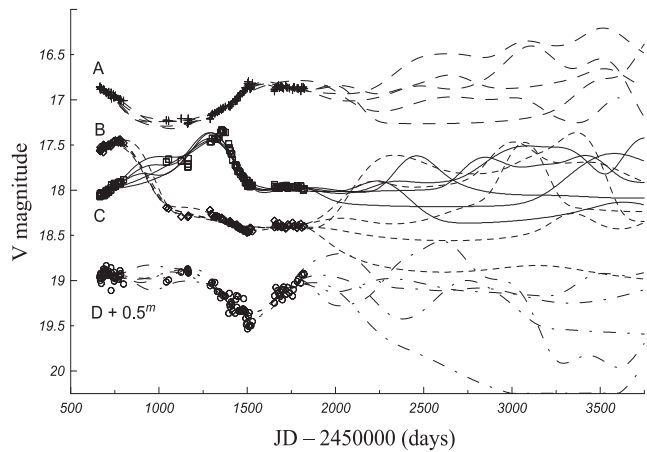}}
 \caption{Some of the most successful simulated light curves plotted against 
 the observed light curves. Microlensing  brightness fluctuations beyond the time
 interval of fitting are approximately of the amplitude and duration observed.}
 \label{fig4}
 \end{figure}

In Fig. 4, some of the most successful simulated light curves are shown 
together with the corresponding observed light curves reduced to their 
zero level. (The quasar light variations have been subtracted as described 
previously). It should be noted that the simulated light curves reproduce 
the observed ones well enough within the time interval of the fitting, and 
no unacceptably large brightness fluctuations are observed outside this 
interval, unlike the results in Fig. 10 of Kochanek (2004).

\section{Conclusions}

In our Q2237 simulations, we have adopted a somewhat simplified version 
of the empirical quasar structure model successfully applied to Q0957 
data by Schild \& Vakulik (2003). We find that our source model consisting 
of a compact inner structure and much larger outer structure with lower 
surface brightness, allows us to 
avoid the effect from the 
standard accretion disc model that large amplitude microlensing brightness 
fluctuations are predicted but not observed, (Jaroszy\'nski et al. 1992, 
Kochanek 2004). We used the method of statistical trials to determine
the inner source dimensions and the contribution of the outer extended 
structure to the total quasar UV-optical luminosity.

Summarizing, we conclude:
\begin{itemize}
\item The proposed source model consisting of two structures, -- an inner
compact structure and an extended outer region, -- provides higher values 
of probability to find "good" simulated light curves as compared to the 
central compact source alone, and produces better fits to long-term light 
curves.

\item We found out that the probability distributions calculated for 
separate image components with two different directions of macroimage 
motion with respect to the shear $\gamma$ are somewhat different.
In principle, this fact might be a clue to determination of the direction 
of the source transverse motion relative to the shear in Q2237+0305. 
Further progress in this important area will require observation of 
many more microlensing events, and many more simulations over longer 
time intervals; this will require greater computational power.
We find that, at the current stage of the investigation, there is no 
possibility to solve the problem, since the maxima of the joint (over 
all components) probability distributions differ insignificantly for the 
two selected directions of the source motion, (see Fig. 3).
 
\item The calculated distributions of the joint probabilities has 
well-marked maxima, and their locations in the domain of parameters 
$\varepsilon$ and $r/r_E$ allow reasonable confidence in their 
determined values which provide the best fit of the simulated light 
curves to the observed ones. The range for the most probable values 
of the relative luminosity $\varepsilon$ of the extended feature, 
determined from probability distributions of the individual 
macroimages, is between 1 and 3, while the estimate of the relative 
size of the compact central feature of the quasar varies within a 
range of  $0.1 < r/r_E < 0.45$. When determined from distributions 
of the joint probabilities, the values of $\varepsilon$ equal 2 in 
both cases, while $r/r_E$ is about 0.4 for A and B motion perpendicular 
to $\gamma$, and 0.15 for A and B moving parallel to the shear $\gamma$.

\item Very significantly, the simulated light curves calculated for 
the proposed two-component source model with the parameters indicated 
above, do not tend to unacceptably increase their amplitudes outside 
the time interval where they were objectively selected according to 
the $\chi^2<\chi_0^2$ criterion, as is seen in Fig. 10 from the work 
by Kochanek (2004).

\item For better comparison, we adopted, following Kochanek (2004), 
a probable projected cosmological transverse source velocity of 
$V_t=3300$ km/s  to determine a linear size of the compact central 
source of $r\approx 2\cdot10^{15}$ cm, $(1.2\cdot10^{15}$cm $<r< 2.8 
\cdot 10^{15}$cm). This size was estimated by Kochanek (2004) to be 
between $r\approx1.4\cdot10^{15}h^{-1}$ cm and $4.5\cdot10^{15}
h^{-1}$ cm for the accretion disc model and for the same transverse 
velocity, ($h=100/H_0$, where $H_0$ is the excepted value of the 
Hubble constant). For the relative size of the source of $0.3 r_E$, 
($0.1r_E<r<0.45r_E$), the estimate for the average microlens mass 
is $\langle m\rangle=1.88\cdot10^{-3}h^2M_\odot$, $(3.08\cdot10^{-4}
h^2M_\odot < \langle m\rangle< 3.3\cdot10^{-2}h^2M_\odot$).

\end{itemize}

Thus we conclude that the proposed two-component model of the Q2237+0305 
quasar structure provides better fit to the actual light curves of the 
four lensed quasar images as compared to the case of a compact central 
source alone. The importance of the extended structure for microlensing 
simulations was suspected for the first time more than ten years ago, 
(Jaroszy\'nski et al. 1992, Witt \& Mao 1994), and was successfully 
demonstrated for Q0957+561 by Schild \& Vakulik (2003). Recently, the 
analysis of the X-ray and optical flux ratio anomalies in ten quadruply 
lensed quasars has made Pooley et al. (2007)  deduce, that the optical 
radiation comes from a region much larger than that expected from the 
thin disk model by a factor of up to 30. Thus quasar microlensing studies 
at X-ray and optical wavelengths are converging to show that standard 
accretion disc models must be supplemented with an extended outer 
structure, contributing a noticeable fraction of the UV-optical 
continuum, and the present report shows how the comparison of simulations
to microlensing observations demonstrates good fit and produces a 
determination of the size of the luminous inner quasar structure 
and its contribution to the total emission.

\section*{Acknowledgments}

This work has been supported by the STCU grant U127. The authors are 
grateful to the unknown referee for very useful critical remarks and 
suggestions.

\label{lastpage}
\end{document}